\documentclass[times]{anzsauth}

\usepackage{moreverb}
\usepackage{enumerate}

\usepackage[colorlinks,bookmarksopen,bookmarksnumbered,citecolor=red,urlcolor=red]{hyperref}

\newtheorem{Remark}{Remark}
\newtheorem{Result}{Result}

\newcommand\BibTeX{{\rmfamily B\kern-.05em \textsc{i\kern-.025em b}\kern-.08em
T\kern-.1667em\lower.7ex\hbox{E}\kern-.125emX}}

\def\volumeyear{2016}

\begin{document}

\runningheads{GEE for vector regression}{A.~HUANG}

\title{On generalized estimating equations for vector regression}

\author{A.~HUANG}

\affiliation{University of Queensland}

\address{School of Mathematics and Physics, The University of Queensland, St Lucia, QLD, Australia, 4072}

\begin{abstract}
Generalized estimating equations \citep[GEE;][]{LZ1986} 
for regression problems with vector-valued responses are examined. When the response vectors are of mixed type (e.g. continuous--binary response pairs), the GEE approach is a semiparametric alternative to full-likelihood copula methods, and is closely related to the mean-covariance estimation equations approach of \citet{PZ1991}. When the response vectors are of the same type (e.g. measurements on left and right eyes), the GEE approach can be viewed as a ``plug-in" to existing methods, such as the \texttt{vglm} function from the state-of-the-art \texttt{VGAM} R package of \citet{Yee2015}. In either scenario, the GEE approach offers asymptotically correct inferences on model parameters regardless of whether the working variance-covariance model is correctly or incorrectly specified. %For testing compound hypothesis on the regression parameters, a simple $\mathcal{F}$-test is developed, with its 
The finite-sample performance of the method is assessed using simulation studies based on a burn injury dataset \citep{Song2007} 
and a sorbinil eye trial dataset \citep{RGL2006}. 
The method is applied to data analysis examples using the same two datasets, as well as on a trivariate binary dataset on three plant species in the Hunua ranges of Auckland \citep{Yee2016}.
\end{abstract}


\ack{The author thanks Thomas Yee, Paul. J. Rathouz, the associate editor and three anonymous referees for comments that improved the paper}

\maketitle

\section{Introduction}
\label{se:intro}

The immense popularity of generalized estimating equations \citep[GEE;][]{LZ1986} for longitudinal data analysis owes much to the fact that it can account for within-cluster correlations without requiring correct specification of the working variance-covariance structure. It achieves this via the much-celebrated sandwich estimator of variance, which is consistent for the true variance matrix of the estimated parameters even when the assumed variance-correlation structure is incorrectly specified. The application of GEEs with a sandwich estimator of variance to general regression problems with vector-valued responses is alluded to in \citet[][Chapter 6]{Song2007}, and is closely related to the mean-covariance estimating equations method of \citet{PZ1991}. This note aims to make the connection between GEEs and vector regression more explicit, as well as to highlight the key differences between some existing and related methods.

Vector regression shares many features with longitudinal data -- both exhibit within-vector correlations and independence across vectors. However, vector regression is distinct from the longitudinal setup in two important ways:

\begin{enumerate}
\item Each vector component may be measurements on different types of variables. In general, we may specify marginal generalized linear models for each component $k$ of a response vector $Y_i$ from subject $i$ via
\begin{eqnarray}
\label{eq:mean}
\mbox{E}(Y_{ik} | X_{ik}) &\equiv & \mu_{ik} = \mu_k (X_{ik}^T\beta_k) \ , \\
\label{eq:var}
\mbox{Var}(Y_{ik}| X_{ik}) &\equiv & \sigma^2_{ik} = \phi_k V_k (\mu_{ik}) \ ,
\end{eqnarray}
for some set of mean functions $\{\mu_k\}$ and some set of variance functions $\{V_k\}$. Here, $k=1,2,\ldots,K$, where $K$ is the number of components in each vector response, and the parameters $\beta_k \in \mathbb{R}^{p_k}$ and $\phi_k$ are the regression coefficients and dispersion parameter, respectively, for each component. The $X_{ik}$s are the corresponding vectors of covariates for each component, which we can collate into a covariate matrix $X_i = (X_{i1}, X_{i2}, \ldots, X_{iK})^T$.
\item The within-vector correlation between two components $Y_{ik}$ and $Y_{ik'}$,
\begin{equation}
\label{eq:corr}
\mbox{Corr}(Y_{ik}, Y_{ik'} | X_i) = \rho_{kk'}(\gamma; \mu_{ik}, \mu_{ik'}) \ ,
\end{equation}
may depend on a vector of correlation parameters $\gamma$ as well as on the marginal means $\mu_{ik}$ and $\mu_{ik'}$. 
% and the covariate vectors $X_{ik}$ and $X_{ik'}$.
\end{enumerate}

\begin{Remark}
In a balanced longitudinal framework, each component is a repeated measure of the same variable, and the mean and variance models (\ref{eq:mean}) and (\ref{eq:var}) typically have the same form for each component. That is,
\begin{eqnarray*}
&& \mu_1(.) \equiv\mu_2(.) \equiv \ldots \equiv \mu_K(.) = \mu(.), \quad \beta_1 \equiv \beta_2 \equiv \ldots \equiv \beta_K = \beta, \\ 
&& V_1(.) \equiv V_2(.) \equiv \ldots \equiv V_K(.) = V(.), \quad \phi_1 \equiv \phi_2 \equiv \ldots \equiv \phi_K = \phi .
\end{eqnarray*}
Moreover, the correlations $\rho_{kk'}$ are typically functions of the correlation parameters $\gamma$ only.
\end{Remark}

%\begin{Remark}
%It is possible for some components to be missing in any given vector. The methods and results in this paper are still valid in this case, provided the data are missing at random (Rubin).
%\end{Remark}

\begin{Remark}
It is possible for some components to share mean parameters. Sometimes this may be due to symmetry arguments, exchangeability, or other special structures in the data. This can be thought of as a hybrid between the general vector regression and longitudinal cases.
\end{Remark}

\begin{Remark}
It is possible for some components to be missing for some observations -- see Section \ref{se:missing}.
\end{Remark}

As is typical in regression problems, the primary interest is in estimating and carrying out joint inferences on the regression parameters $\beta = (\beta_1^T, \beta_2^T,\ldots, \beta_K^T)^T$. The variance--correlation structure itself is of secondary importance, coming into consideration when evaluating the sampling variability of parameter estimates. A key feature of the approach examined in this note is that the variance and correlation structures need not be correctly-specified. As in the longitudinal setting of \citet{LZ1986}, we find in Section \ref{se:estimation} that the sandwich estimator of variance is consistent even if the working variances and correlations are misspecified. Of course, the closer the working variance-correlation structure is to the truth, the more efficient and accurate the GEE framework tends to be for finite sample sizes, as demonstrated in the simulation studies in Section \ref{se:sim}.

There are two main existing frameworks for vector regression. The method described in \citet[][Chapter 6]{Song2007} uses copula functions to combine marginal distributions with association structures to construct full probability models for the data. The Gaussian copula is the most popular choice, although other copula functions exist and can also be used. A main drawback of copula-based methods, as with other parametric models, is that the marginal distributions, association structures and copula function are all implicitly assumed to be correctly-specified. 

An alternative to full probability models is an estimating equations approach. The recent book from \citet{Yee2015}, which accompanies the state-of-the-art R package \texttt{VGAM} \citep{Yee2016}, covers marginal models and estimating equations for vector GLMs %and generalized additive models 
that are similar in aims to the current proposal. However, the function \texttt{vglm} for fitting vector generalized linear models currently cannot handle vectors of mixed type, such as continuous-binary pairs. For vectors with components of the same type, the approach examined here can be implemented as a simple ``plug-in" to \texttt{vglm}, offering an adjustment to model-based standard errors in cases where the working model is incorrectly specified. This typically involves just a few lines of code; an example is given in Appendix \ref{ap:plugin}. 

For vector responses of mixed type, an alternative estimating equations approach is the mean-covariance estimating equations (MCEE) framework of \citet{PZ1991} which is motivated from a working quadratic exponential family for the responses. More precisely, the response vectors $Y_i = (Y_{i1}, Y_{i2}, \ldots, Y_{iK})^T$ are assumed to have joint densities of the form
\begin{equation}
\label{eq:quadexp}
\mbox{Pr}_i(y_i) \, \propto \, \exp\left\{ \theta_i^T y_i + \lambda_i^T w_i + c_i(y_i) \right\} \ ,
\end{equation}
where $w_i = (y_{11}^2, y_{11} y_{12}, \ldots, y_{22}^2, y_{23} y_{24}, \ldots)$, $c_i: \mathbb{R}^K \to \mathbb{R}$ are functions that characterises the ``shape" of the joint densities, and $\theta_i = (\theta_{i1}, \theta_{i2}, \ldots, \theta_{iK})^T$ and $\lambda_i = (\lambda_{i1}, \lambda_{i2},\ldots,\lambda_{iK})^T$ are the canonical parameters in the quadratic exponential family which in turn define the marginal means, $\mu_{i1}, \mu_{i2}, \ldots, \mu_{iK}$, and the variance and covariances, $\sigma_{i1}^2, \sigma_{i12}, \ldots, \sigma_{i2}^2, \ldots, \sigma_{iK}^2$.
%$\mu_i = (\mu_1(X_{i1}^T \beta_1), \mu_2(X_{i2}^T \beta_2), \ldots, \mu_K(X_{iK}^T \beta_K))^T$
A full parametric model specification would involve specifying the functions $c_i(.)$, but the MCEE can be formulated without reference to any specific probability distribution, because the score equations for the mean, variance and covariance parameters can be shown to have the same functional form regardless of the underlying quadratic exponential family \citep[see Appendix 2 of][]{PZ1991}. The subsequent estimating equations are given by Equation 6 in \citet{PZ1991}, from which it can be seen that in addition to specifying the marginal means (\ref{eq:mean}), marginal variances (\ref{eq:var}) and within-vector correlations (\ref{eq:corr}), there are additional requirements of a working model for $\mbox{Var}(s_i)$, the variance  of the empirical covariance vectors $s_{i} = (s_{i11}, s_{i12}, \ldots, s_{i KK})^T$, where $s_{i k k'} = (Y_{ik} - \mu_{ik})(Y_{ik'} - \mu_{ik'})$ for $k, k' = 1,2,\ldots, K$, and a working model for $\mbox{Cov}(Y_i, s_i)$, the covariance between the response vectors $Y_i$ and $s_{i}$ \citep[see Equation 7 of][]{PZ1991}. Apart from some special cases, it is generally difficult to specify appropriate models for these additional higher-order terms. 

The MCEE approach is useful when there is direct interest in the variance-covariance structure. If the interest lies primarily in the regression parameters, then the introduction of a second-order model via the quadratic exponential family (\ref{eq:quadexp}) adds an extra level of model complexity that is not required. Moreover, the quadratic exponential family model is typically inappropriate if some of the components are non-Gaussian. For example, if one of the components is binary, then the mean completely determines the variance and the corresponding quadratic terms need to be eliminated from the quadratic exponential family (\ref{eq:quadexp}) and subsequent estimating equations in order to avoid degeneracy \citep[see Section 4 of][]{PZ1991}. If one of the components is a gamma-type random variable, say, then it is perhaps more sensible to consider a second-order exponential family defined in terms of the sufficient statistics $(y, \log(y))$ rather than $(y, y^2)$. The latter is appropriate for approximately Gaussian responses.

In contrast, the GEE approach we examine here is simpler to specify, and can be formulated without reference to any underlying joint distribution for the data -- one only needs to specify a set of marginal mean models (\ref{eq:mean}), a set of working variance functions (\ref{eq:var}), and a set of working within-vector correlation functions (\ref{eq:corr}). Of course, context-specific parametric models, such as the hierarchical continuous-binary model of \citet{FL1995}, the joint continuous-discrete models of \citet[][Chapter 24]{MV2005} or the latent variable models of \citet{Dunson2000}, can also be used to motivate the working variances (\ref{eq:var}) and correlations (\ref{eq:corr}), but the GEE approach can be formulated without actually imposing any of these underlying joint distributions. The sandwich estimator of variance can then be applied for asymptotically correct inferences on the mean model (\ref{eq:mean}) even under misspecification of the working variance and/or correlation model.

%{\color{magenta} In Section 2, we ... Section 3 outlines the estimation procedure and states the main results. Two sets of simulations are used in Section 4 to investigate the finite-sample performance of the approach under model misspecification. An analysis of the original two datasets are given in Section 5. The discussion in Section 7 highlights current and future developments in small-sample adjustments and software implementation.}

\section{Two motivating examples}
\label{se:examples}
We highlight the differences between the copula, GEE and MCEE approaches using the following two examples. The first is a dataset with vector responses of mixed type, and the second is a dataset with bivariate responses on a finite lattice. It is difficult to specify joint distributions for the data in either scenario, and it is in such cases where a GEE approach proves invaluable. 

In the following, we drop the first subscript $i$ in our notation and simply write $Y_k$ for the $k$th component of a generic response vector $Y$, and $X_k$ for the corresponding row of a generic design matrix $X$, for clarity of exposition. The full dataset can then be considered as $n$ independent copies of the generic predictor-response pair $(X,Y)$.

\subsection{Burn injury data} 
\label{se:ex1}
\citet{Song2007} describes a dataset from \citet{FG1996} containing 981 cases of burn injuries, with two response variables of mixed type. These are severity of burns as measured by $Y_{1} = \log(\mbox{burn area} + 1)$ and disposition of death as measured by $Y_{2} = 1$ for death and 0 for survival. It is of interest to see whether severity of burns and probability of death is related to age.

Burn severity is continuous and disposition of death is binary, so a default set of marginal mean functions is
$$
E(Y_{1}| {\rm age}) = \mu_{1} = \beta_{10} + \beta_{11} {\rm age}\ , \quad E(Y_{2}|{\rm age}) = \mu_{2} = \frac{ \exp(\beta_{20} + \beta_{21} {\rm age}) }{1+\exp(\beta_{20} + \beta_{21} {\rm age}) } \ ,
$$
with a corresponding set of marginal variance functions,
$$
\mbox{Var}(Y_{1}|{\rm age}) = \sigma^2 \ ,\quad \mbox{Var}(Y_{2}|{\rm age}) = \mu_{2} (1-\mu_{2})  \ .
$$
Here, the dispersion parameters are taken to be $\phi_1 = \sigma^2$ and $\phi_2 \equiv 1$, respectively.

For a copula-based approach, the marginal mean and variance functions are first formalized into marginal distributions, say,
$$
Y_1|{\rm age} \sim N(\mu_1, \sigma^2) \ , \quad Y_2|{\rm age} \sim \mbox{Bernoulli}(p=\mu_2) \ .
$$
Then, a joint distribution for the response pair can be constructed using a Gaussian copula, for example, via
$$
F(y_1, y_2) = G\{\Phi^{-1}(F_1(y_1)), \Phi^{-1}(F_2(y_2)) \} \ .
$$
Here, $F_1$ is the cumulative distribution function (cdf) of a $N(\mu_1, \sigma^2)$, $F_2$ is the cdf of a Bernoulli$(p=\mu_2)$, $\Phi$ is cdf of a standard normal, and $G$ is the cdf of a bivariate normal with mean 0, unit variances and correlation $\rho$, which describes the level of association between burn severity and incidence of death. Other copula functions can also be used. The model parameters are then estimated via maximum likelihood, but the procedure is typically computationally demanding as second order derivatives of the log-likelihood are not readily available  \citep[see][Section 6.4]{Song2007}. Alternative computational methods include the Maximization-by-parts approach from \citet{SFK2005} and the Gauss-Newton type approach of \citet{Ruppert2005}. The R package \texttt{gcmr} of \citet{MV2015} implements Gaussian copula regression models via importance sampling, but currently cannot handle data of mixed type.
%Subsequent inferences on model parameters based on the log-likelihood function assume that the marginal distributions and copula function are correctly specified, and no adjustments are readily available in case of model misspecification for mixed responses.

In contrast, a GEE approach simply requires specifying a working correlation model for the data. This can be motivated from a working joint distribution, such as that induced by a copula approach, or specified without reference to any joint distribution. For example, 
%a naive assumption here would be that the severity of burns is uncorrelated with death status within each individual,
%$$
%\mbox{Corr}(Y_{1},Y_{2} |age) = 0 \ ,
%$$
%or that there is some general correlation,
a simple model here would be that severity of burns has an arbitrary constant correlation with death status within each individual,
$$
\mbox{Corr}(Y_{1},Y_{2} | {\rm age}) = \rho \ , \mbox{ for some } \rho \in (-1,1) \ .
$$
A more flexible model is to allow the correlation $\rho = \rho(\gamma; \mu_1, \mu_2)$ to change with the marginal means. As we show in Section \ref{se:estimation}, any reasonable working variance and correlation model can be used, with model misspecifications being adjusted for using a sandwich estimator of variance.

For a MCEE model, in addition to a working correlation model, one needs to specify working models for $\mbox{Var}(s)$ and $\mbox{Cov}(Y, s)$, where $s = (s_{11}, s_{12})^T = ((Y_1-\mu_1)^2, (Y_1 - \mu_1)(Y_2- \mu_2))^T $. Note that $s_{22} = (Y_2 - \mu_2)^2$ has been removed from the empirical covariances as the mean of a binary response determines its variance. \citet{PZ1991} offer three ways of specifying these additional terms: 
\begin{enumerate}[i]
\item Independence working matrices, with $\mbox{Cov}(Y, s)=0$ and
$$
\mbox{Var}(s) = \begin{pmatrix}
2 \sigma^4 & 0 \\
 & \sigma^2 \mu_2(1-\mu_2)  
% &  & 2 \mu_2^2(1-\mu_2)^2
\end{pmatrix} \ ;
$$ 
\item Gaussian working matrices, with $\mbox{Cov}(Y, s)=0$ and 
$$
\mbox{Var}(s) = \begin{pmatrix}
2 \sigma^4 & 2 \rho \sigma^3 \sqrt{\mu_2 (1-\mu_2)}  \\ %&  2 \rho^2 \sigma^2 \mu_2 (1-\mu_2) \\
 & (1+\rho^2) \sigma^2 \mu_2(1-\mu_2)  %& 2\rho \sigma \mu_2^{3/2} (1-\mu_2)^{3/2}  \\
% &  & 2 \mu_2^2(1-\mu_2) ^2
\end{pmatrix} \ ;
$$ 
\item Gaussian matrices with common third and fourth-order correlations, with
$$
\mbox{Cov}(Y,s) = \begin{pmatrix}
\gamma_{111} \sigma^2 \sqrt{\mu_2(1-\mu_2)} & \gamma_{112} \sigma^2 \sqrt{\mu_2(1-\mu_2)} \\ %& \gamma_{122} \sigma \mu_2(1-\mu_2) \\
 & \gamma_{212} \sigma \mu_2(1-\mu_2) % & \gamma_{222} \mu_2 ^{3/2} (1-\mu_2)^{3/2}
\end{pmatrix}
$$ 
and
$$
\mbox{Var}(s) = \begin{pmatrix}
(2+\delta_{1111}) \sigma^4 & (2\rho + \delta_{1112}) \sigma^3 \sqrt{\mu_2(1-\mu_2)} \\ % & (2\rho + \delta_{1122}) \sigma^2 \mu_2(1-\mu_2) \\
 & (1 + \rho^2 + \delta_{1212}) \sigma^2 \mu_2 (1-\mu_2) % & (2\rho + \delta_{1222}) \sigma \mu_2^{3/2} (1-\mu_2)^{3/2} \\
%  & & (2 + \delta_{2222}) \mu_2^2 (1-\mu_2)^2
\end{pmatrix} \ ,
$$ 
where $\gamma_{111}, \gamma_{112}, \gamma_{212}$ and $\delta_{1111}, \delta_{1112}, \delta_{1212}$ are additional parameters to be estimated. \citet{PZ1991} prescribe a method-of-moments approach for estimating these parameters that are $\sqrt{n}$-consistent.
\end{enumerate}
As mentioned in the introduction, the MCEE approach is useful when the variance-covariance structure is directly of interest. If the interest lies primarily in the regression parameters, then the introduction of these additional terms adds an extra level of model complexity over the GEE approach that is not required.

\subsection{Sorbinil Retinopathy Trial data}
\label{se:ex2}
\citet{RGL2006} describes a dataset consisting of pairs of itching scores $(Y_L, Y_R)$ from left and right eyes of 41 subjects in a Sorbinil Retinopathy Trial. The scores range from 0 (no itch at all) to 4 (severe incapacitating itch) in increments of 0.5. The active treatment was sorbinil, which could be applied to both eyes, just the left eye, just the right eye, or to neither eye. The data are displayed in Table \ref{tab:1} of Appendix \ref{se:appA}.

\begin{table}[H]
\footnotesize
\centering
\caption{Sorbinil trial data -- itching scores in left and right eyes by treatment group, from \citet{RGL2006}}
\begin{tabular}{ccccccccccc}
\toprule
 \multicolumn{11}{c}{Treatment combination} \\
\midrule
sorbinil & sorbinil & \hspace{2mm} & sorbinil & placebo & \hspace{2mm} & placebo & sorbinil & \hspace{2mm} & placebo & placebo \\
\cline{1-2} \cline{4-5} \cline{7-8} \cline{10-11}
left & right & & left & right & & left & right & & left & right \\
\midrule
2.0 & 2.0 & & 1.0 & 1.5 & & 2.5 & 2.0 & & 3.0 & 3.0 \\
1.0 & 1.0 & & 2.0 & 2.5 & & 2.5 & 2.5 & & 2.0 & 3.0 \\
0.5 & 2.0 & & 3.0 & 1.0 & & 3.0 & 3.0 & & 2.5 & 2.5 \\
2.5 & 1.0 & & 2.0 & 3.0 & & 2.5 & 2.0 & & 1.0 & 3.0 \\
3.0 & 2.5 & & 3.0 & 2.5 & & 1.0 & 0.5 & & 2.0 & 2.5 \\
2.0 & 2.5 & & 2.0 & 3.0 & & 2.0 & 0.0 & & 2.0 & 1.0 \\
     &       & & 3.0 & 3.0 & & 3.0 & 2.5 & & 2.0 & 2.0 \\
     &       & & 0.5 & 1.5 & & 3.0 & 1.0 & &       &      \\
     &       & & 3.0 & 3.0 & & 2.0 & 1.5 & &       &      \\
     &       & & 3.0 & 3.0 & & 0.5 & 0.0 & &       &      \\
     &       & & 3.0 & 3.0 & & 2.5 & 1.5 & &       &      \\
     &       & & 1.0 & 2.0 & & 2.0 & 2.0 & &       &      \\
     &       & & 1.0 & 2.0 & & 2.5 & 2.5 & &       &      \\    
     &       & & 1.5 & 2.5 & & 2.5 & 2.5 & &       &      \\
\bottomrule
\end{tabular}
\label{tab:1}
\end{table}

Both $Y_L$ and $Y_R$ are scores between 0 and 4 inclusive, so the mean functions should also be constrained in this range for any value of the covariate(s). Moreover, it is sensible for the variance functions to satisfy
$\mbox{Var}(Y_k) \to 0$ as $E(Y_k) \to 0 \mbox{ or } E(Y_k) \to 4$ for either $k=L, R$. In particular, constant variance is not appropriate here. 

One way to take these constraints into consideration is to transform the itchiness scores into the unit interval $[0,1]$ via $\tilde Y_L = Y_L/4$  and $\tilde Y_R = Y_R/4$, and then treat the transformed responses as pseudo-proportions \citep[c.f. example 9.2.4 in][]{MN1989} using a pair of logistic mean functions,
\begin{eqnarray*}
E(\tilde Y_L|\mbox{ treatment} ) &=& \mu_L = \frac{\exp \left[\beta_{L0} + \beta_{L1} I(\mbox{sorbinil}_L =1) \right]}{1+\exp\left[\beta_{L0} + \beta_{L1} I(\mbox{sorbinil}_L =1) \right] } \ , \\
E(\tilde Y_R|\mbox{ treatment} ) &=& \mu_R = \frac{\exp \left[\beta_{R0} + \beta_{R1} I(\mbox{sorbinil}_R =1)\right]}{1+\exp \left[\beta_{R0} +\beta_{R1} I(\mbox{sorbinil}_R =1)\right]} \ ,
\end{eqnarray*} 
where $I(\mbox{sorbinil}_L =1)$ is an indicator for sorbinil treatment in the left eye, and similarly for $I(\mbox{sorbinil}_R =1)$. A corresponding pair of working variance functions is
\begin{eqnarray*}
\mbox{Var}(\tilde Y_L| \mbox{treatment}) &=& \phi \, \mu_L \left( 1- \mu_L \right)\\
\mbox{Var}(\tilde Y_R | \mbox{treatment}) &=& \phi \, \mu_R \left( 1- \mu_R \right) \ .
\end{eqnarray*}
Here, we have assumed equal dispersion in the two eyes, so that the dispersion parameters are $\phi_1 = \phi_2 = \phi$.

For a copula-based approach, the marginal mean models and variance functions need to be formalized into marginal distributions. However, it is not easy to specify distributions on the finite set of values $\{0, 0.125, 0.25, \ldots, 0.875, 1\}$, or on the original scale $\{0, 0.5,1,\ldots, 3.5, 4\}$, that satisfy the marginal mean and variance constraints. Thus, a semiparametric estimating equations approach proves to be invaluable here, as one only needs to specify a working correlation model without necessarily invoking marginal or joint distributions for the data.

For a GEE approach, a simple working model for the within-subject correlation between the two eyes is
\begin{eqnarray*}
\mbox{Corr}(\tilde Y_L, \tilde Y_R | \mbox{treatment}) &=& \rho \ , \mbox { for some } \rho \in (-1, 1) \ ,
\end{eqnarray*}
or perhaps a more flexible model $\rho = \rho(\gamma; \mu_L, \mu_R)$ that allows the correlation to change with the marginal means, such as that induced by the odds ratio model of \citet{Yee2015}. This is considered in more detail in Section \ref{se:sim2}.
%$$
%\log \left\{ \frac{P(Y_L =1, Y_R =1) P(Y_L = 0, Y_R = 0)}{P(Y_L=1, Y_R = 0) P(Y_L=0, Y_R=1)}\right\} = \gamma
%$$
Again, any reasonable working model can be used here, with potential misspecifications being adjusted for using a sandwich estimator of variance.

To formulate a MCEE model, in addition to a working correlation model one needs to specify working models for $\mbox{Var}(s)$ and $\mbox{Cov}(\tilde Y, s_{LR})$, where $s_{LR} =(\tilde Y_L - \mu_L)(\tilde Y_R- \mu_R)$. Note that $s_{LL} = (\tilde Y_L - \mu_L)^2$ and $s_{RR} = (\tilde Y_R - \mu_R)^2$ have been removed from the empirical covariances because the variances of proportions are determined by their means. The following are three specifications described in \citet{ZP1990}:
\begin{enumerate}[i]
\item Independence, with $\mbox{Cov}(\tilde Y, s_{LR})=0$ and $\mbox{Var}(s_{LR}) \propto \mu_L \mu_R (1-\mu_L) (1-\mu_R) $.
\item Independence with structural nonzeros, with  
\begin{eqnarray*}
\mbox{Cov}(\tilde Y_L, s_{LR}) &\propto& (1-2\mu_L) \rho \sqrt{\mu_L\mu_R(1-\mu_L)(1-\mu_R)} \ ,\\
\mbox{Cov}(\tilde Y_R, s_{LR}) &\propto& (1-2\mu_R) \rho \sqrt{\mu_L\mu_R(1-\mu_L)(1-\mu_R)}  \ ,\\
\mbox{Var}(s_{LR}) &\propto& (1 - \rho^2) \mu_L \mu_R (1-\mu_L) (1-\mu_R) \\ 
& & + (1-2\mu_L)(1-2\mu_R) \rho \sqrt{\mu_L \mu_R (1-\mu_L) (1-\mu_R)} \ .
\end{eqnarray*}
%\item Common third and fourth order correlations, with
%\begin{eqnarray*}
%\mbox{Cov}(\tilde Y_L, s_{LR}) &=& \gamma_{LLR} \, \mu_L(1-\mu_L) \sqrt{\mu_R(1-\mu_R)} \ ,\\
%\mbox{Cov}(\tilde Y_R, s_{LR}) &=& \gamma_{RLR} \, \mu_R(1-\mu_R) \sqrt{\mu_L(1-\mu_L)} \ ,\\
%\mbox{Var}(s_{LR}) &=&
%\end{eqnarray*}
\item Gaussian scores, with $\mbox{Cov}(\tilde Y, s_{LR}) = 0$ and $\mbox{Var} \propto (1+\rho^2) \mu_L \mu_R(1-\mu_L)(1-\mu_R)$.
\end{enumerate}

The primary interest here is to estimate and make inferences on the treatment effect of Sorbinil on reducing eye irritation. However, we may also be interested in testing for {\it symmetry} in the two eyes, and whether treatment applied to one eye {\it interferes} with itchiness levels of the other eye. The GEE approach enables us to make valid inferences on the regression parameters without having to specify correct marginal variances, within-subject correlation, and working models for higher-order moments. In addition to being simpler to formulate, the GEE approach is also straightforward to implement. Section \ref{se:estimation} outlines the estimation and inference procedure in the GEE vector regression framework.

\section{Estimation and inference}
\label{se:estimation}
Given a set of variance functions (\ref{eq:var}) and correlation functions (\ref{eq:corr}), a working variance-covariance matrix $W_i \equiv W_i(\beta; \gamma; \phi)$ for each observation $i$ can be constructed via
$$
W_i = \Sigma_i^{1/2} R_i \Sigma_i^{1/2} \ ,
$$
where $\Sigma_i = \mbox{diag}(\sigma^2_{i1}, \sigma^2_{i2}, \ldots, \sigma^2_{iK})$ is a diagonal matrix of the variances of each component of $(Y_{i1}, Y_{i2}, \ldots, Y_{iK})^T$, and $R_i$ is the correlation matrix induced by (\ref{eq:corr}). Note that, unlike the longitudinal case, dispersion parameters are generally included in the working covariance matrices $W_i$. This is because each component may have different dispersion parameters which cannot be factorised out. If a working independence correlation structure is used, however, then the dispersion parameters drop out, as in the longitudinal setting.

An estimator of $\beta = (\beta_1^T, \beta_2^T,\ldots, \beta_K^T)^T$ can be defined as the solution to the set of generalized estimating equations, 
$$
0 = \sum_{i=1}^n D_i^T W_i^{-1} S_i 
$$
where $S_i \equiv S_i(\beta) = (Y_{i1} - \mu_{i1}, Y_{i2} - \mu_{i2}, \ldots, Y_{iK} - \mu_{iK})^T$ are the residuals and $D_i \equiv D_i(\beta) = \{ \partial \mu_{ik}/\partial \beta_{k'} \}_{k,k' = 1,2,\ldots, K}$ is a matrix of derivatives. We iteratively solve for the regression coefficients and the correlation and dispersion parameters by alternating the following two steps. 

Given current estimates $\hat \beta$, $\hat \phi$ and $\hat \gamma$, an update for $\beta$ can be obtained via Fisher scoring \citep[c.f. equation 8 in][]{LZ1986}
$$
\hat \beta^{(1)} = \hat \beta - \left\{\sum_{i=1}^n D_i^T(\hat \beta) W_i^{-1}(\hat \beta; \hat \gamma; \hat \phi) D_i (\hat \beta) \right\}^{-1} \left\{ \sum_{i=1}^n D_i^T(\hat \beta) W_i^{-1}(\hat \beta; \hat \gamma; \hat \phi) S_i(\hat \beta) \right\} \ .
$$
Conversely, given an estimate of $\beta$, consistent estimates of the dispersion $\phi$ and correlation $\gamma$ parameters can be obtained via method-of-moment approaches analogous to those from Section 4 of \citet{LZ1986}. For example, in the general case where each component $k$ has its own dispersion parameter $\phi_k$, we can estimate $\phi_k$ via 
$$
\hat \phi_k^{-1} = \frac{1}{n-p_k} \sum_{i=1}^n \frac{(Y_{ik} - \hat \mu_{ik})^2} {V_k(\hat \mu_{ik})} \ , \quad \ k=1,2,\ldots, K \ .
$$
For an unstructured working correlation model, $R(\alpha) = \{\alpha_{kk'} \}_{k,k'=1,2,\ldots, K}$ can be estimated via
$$
\hat R = \frac{\hat \phi}{n} \sum_{i=1}^n \Sigma_i^{-1/2} S_i S_i^T \Sigma_i^{-1/2} \ .
$$
Note that other popular correlation models from the longitudinal setting, such as exchangeable, $m$-dependence, auto-regressive and band diagonal structures, are generally not applicable to vector regression because the indices of the vector components are exchangeable labels. Thus, the working independence and unstructured correlation models are perhaps the only two generally applicable working correlation structures for vector regression. Other dependence models, such as the odds ratio model for bivariate binary regression of \citet{Yee2015}, will be specific to the problem at hand.

% \section{Main results}
The following results are vector regression analogues of the results in  \citet{LZ1986}, concerning the asymptotic normality of the GEE estimator and consistency of the sandwich estimator for the asymptotic variance.

\begin{Result}[(Asymptotic normality)]
\label{result:1}
Under conditions of Theorem 2 from \citet{LZ1986}, we have
$
\sqrt{n} (\hat \beta - \beta^*) \to N(0, V_\beta),
$
where 
$$
V_\beta = \lim_{n \to \infty} n \left(\sum_{i=1}^n D_i^T W_i^{-1} D_i \right)^{-1} \left\{\sum_{i=1}^n D_i^T W_i^{-1} \mbox{cov}(Y_i) W_i^{-1} D_i \right\} \left(\sum_{i=1}^n D_i^T W_i^{-1} D_i \right)^{-1} \ .
$$
\end{Result}
\begin{Result}[(Sandwich estimator of variance)]
\label{result:2}
$V_\beta$ can be consistently estimated by
$$
\hat V_\beta = n \left(\sum_{i=1}^n \hat D_i^T \hat W_i^{-1} \hat D_i \right)^{-1} \left\{\sum_{i=1}^n \hat D_i^T \hat W_i^{-1} \hat S_i \hat S_i^T \hat W_i^{-1} \hat D_i \right\} \left(\sum_{i=1}^n \hat D_i^T \hat W_i^{-1} \hat D_i \right)^{-1} \ ,
$$
where $\hat D_i = D_i(\hat \beta)$, $\hat S_i = S_i(\hat \beta)$, and $\hat W_i = W_i(\hat \beta; \hat \gamma; \hat \phi)$ are the plug-in estimators.
\end{Result}

In Result \ref{result:2}, the term $\left(\sum_{i=1}^n \hat D_i^T \hat W_i^{-1} \hat D_i/n \right)^{-1}$ on the outside is the unadjusted, or model-based, estimator of variance. It is consistent for the asymptotic variance only if the working variance--covariance model is correct. In contrast, the sandwich estimator of variance $\hat V_\beta$ is consistent regardless of the true variance--covariance structure. Underlying this, of course, is the assumption that the mean model (\ref{eq:mean}) is correct.

Results \ref{result:1} and \ref{result:2} can then be combined to carry out joint inferences on regression coefficients, as prescribed in Result \ref{result:3} below.

\begin{Result}[(Joint inferences)] 
\label{result:3}
Under the null hypothesis $H_0: M \beta = \delta$, where $M$ is a given $r \times p$ matrix of full rank and $\delta$ a given vector, the quadratic form
$$
F = \frac{n}{r}  (M \hat \beta - \delta)^T \left(M \hat V_\beta M^T \right)^{-1} (M\hat \beta - \delta) 
$$
has an approximate $\mathcal{F}_{r, n-p}$ distribution for large $n$, where $p=p_1 + p_2 + \ldots + p_K$ is the total number of regression parameters.

\end{Result}

\section{Simulations}
\label{se:sim}
To assess the finite-sample accuracy of our asymptotic results, particularly under model misspecification, we carry out two sets of simulations. The first is based on the continuous-binary burn injury dataset from Example \ref{se:ex1}, and focuses on the accuracy of the sandwich estimator $\hat V_\beta$ and its inverse $\hat V_\beta^{-1}$ for estimating the variance and precision matrices, respectively, of the GEE estimator $\hat \beta$. The second is based on the bivariate Sorbinil eye trial dataset from Example \ref{se:ex2}, and focuses on comparing the GEE method with the copula approach under different working correlation models. In both settings, the Type I errors for testing model parameters using the $\mathcal{F}$-test of Result \ref{result:3}, with and without sandwich adjustments, are compared to their nominal levels.
\subsection{Burn injury simulations}
%To assess the finite-sample performance of our asymptotic results under model misspecification, 

We simulate from the hierarchical continuous-binary model of \citet{FL1995} to emulate the burn injury data from Example \ref{se:ex1}. More precisely, to generate each dataset we first sample $n=200, 400$ or $800$ ages from the 981 original observations, then, conditional on the ages, we generate burn severity $Y_1$ and incidence of death $Y_2$ via the hierarchical model,
\begin{eqnarray*}
Y_2|\mbox{age} &\sim & \mbox{Bernoulli}\{\mu_2\} \ , \mbox{ where } \mu_2 = \frac{\exp(-4.0521 + 0.0527\mbox{ age})}{1+\exp(-4.0521 + 0.0527\mbox{ age})} \ , \\
Y_1| \mbox{age}, Y_2 &\sim & N\left\{6.6980 + 0.0039 \mbox{ age} +\gamma(Y_2 - \mu_2), \  1.26^2 \right\} \ .
\end{eqnarray*}
By generating $Y_1$ conditional on $Y_2$, we induce within-vector correlation for each subject, with the parameter $\gamma$ controlling the strength of this correlation. Here, we set $\gamma = 5$. The parameter values are taken from the fitted model to the original dataset, so as to emulate the original values. We repeat this procedure to generate 1,000 datasets. 

For each synthetic dataset, we fit the GEE model from Example \ref{se:ex1} with the working independence assumption, computing the GEE estimate of $\beta$ along with its estimated variance matrix using the sandwich estimator from Result \ref{result:2}. Note that the marginal variance function for $Y_1$ is misspecified, along with the working independence assumption. It is in such cases where a robust adjustment to the variances might be invaluable. The code for computing the sandwich adjustment is particularly simple and provided in Appendix \ref{ap:B1}. In the special case of working independence, the GEE model coincides with an independent copula model with Gaussian and binary marginal distributions.

%using the pair of marginal mean models,
%\begin{eqnarray*}
%E(Y_1|\mbox{age}) &=& \mu_1 = \beta_{10} + \beta_{11} \mbox{age}\ ,\\
%E(Y_2|\mbox{age}) &=& \mu_2 = \frac{ \exp(\beta_{20} + \beta_{21} \mbox{age}) }{1+\exp(\beta_{20} + \beta_{21} \mbox{age}) } \ ,
%\end{eqnarray*} 
%a corresponding pair of working variance functions,
%\begin{eqnarray*}
%\mbox{Var}(Y_1|\mbox{age}) &=& \sigma^2 \ , \\
%\mbox{Var}(Y_2|\mbox{age}) &=& \mu_2 (1-\mu_2)  \ .
%\end{eqnarray*}
%and a working independence correlation model,
%$$
%\mbox{Corr}(Y_1, Y_2| \mbox{age}) = 0 \ .
%$$

To assess the accuracy of the sandwich variance estimator $\hat V_\beta$ prescribed in Result \ref{result:2}, we compute the matrix 1-norm $
\|\hat V_\beta - V_\beta\|_1 \ = \sup \{ \|\hat V_\beta u - V_\beta u\|_1: \|u\| = 1 \}
$
for each simulation, where $V_\beta$ is the empirical ``true" variance-covariance matrix across our simulations. We also examine the  estimation accuracy for the precision matrix, $V_\beta^{-1}$, which is arguably more pertinent for inferences than the variance. For each simulation, we compute the analogous matrix norm $\| \hat V_\beta^{-1} - V_\beta^{-1} \|_1$. The results are summarised in rows 11--12 of Table \ref{tab:2}.

We see that the sandwich adjustment offers much improved performance in estimating both the variance and precision matrices, with the improvement becoming more pronounced when the sample size increases. In contrast, the unadjusted precision estimator actually diverges with increasing sample size, which leads to biased inferences on the regression parameters. 

Indeed, the Type 1 errors (rows 13--15 in Table \ref{tab:2}) for jointly testing the true parameter values $H_0: \beta_{10} = 6.6980, \beta_{11} = 0.0039, \beta_{20} = -4.0521$ and $\beta_{21} = 0.0527$ at nominal 1\%, 5\% and 10\% levels are seen to be severely biased when the unadjusted variances are used. On the other hand, inferences using the adjusted variances have Type I error rates that are much closer to their nominal levels. These tests were carried out using the $\mathcal{F}$-test from Result \ref{result:3}, calibrated against the $\mathcal{F}$-distribution with 4 and $n-4$ degrees of freedoms.

Table \ref{tab:2} also displays the accuracy of the adjusted and unadjusted variance estimators at a component-wise level. More precisely, the adjusted and unadjusted means and standard deviations of the estimated standard deviations $\hat{\mbox{sd}}(\hat \beta_{kj})$ and correlations $\hat{\mbox{corr}}(\hat \beta_{kj}, \hat \beta_{k'j'})$ of $\hat \beta$ are given in rows 1--10. We see that the adjusted estimator leads to more precise estimation of each standard deviation and correlation term.

%\begin{table}[H]
%\footnotesize
%\caption{Burn injury simulations -- mean accuracy of the sandwich estimator for the variance  $(V_\beta)$ and precision $(V^{-1}_\beta)$ matrices, and Type I errors for jointly testing all model parameters at nominal 1\%, 5\% and 10\% levels. $N=1000$ simulations each with sample sizes of $n=200,400$ and $800$.}
%\centering
%\begin{tabular}{cccrrrrrrrrr}
%\toprule
% & &  & \multicolumn{2}{c}{\underline{\hspace{6mm} accuracy \hspace{6mm} }} & & \multicolumn{3}{c}{\underline{\hspace{2mm}Type I errors \hspace{2mm}}}\\
%$n$ & & method & variance & precision & & 1\% & 5\% & 10\%\\ 
%\midrule
%$200$ & & unadjusted & 40.6 & 1425.1  & & 3.8 & 11.3 & 17.1 \\
%& & adjusted & 28.3 & 418.1 & & 1.9 & 7.2 & 13.2 \\
%$400$ & & unadjusted & 30.7 & 1555.0 & & 3.7 & 10.1 & 15.4 \\
%& & adjusted & 17.7 & 281.4 & & 1.4 & 6.0 & 11.3
%\\
%$800$ & & unadjusted & 26.2 & 1573.6 & & 3.1 & 9.1 & 13.4 \\
%& & adjusted & 12.5 & 194.4 & & 0.9 & 5.1 & 9.7 \\
%\bottomrule
%\end{tabular}
%\end{table}

\begin{table}
\footnotesize
\caption{Burn injury simulations -- (i) component-wise means (and standard deviations) of estimated asymptotic standard deviations and correlations, (ii) mean accuracy of estimated variance and precision matrices, and (iii) Type I errors for jointly testing all model parameters at nominal 1\%, 5\% and 10\% levels. $N=1000$ simulations each with sample sizes $n=200, 400$ and $800$.}
\centering
\begin{tabular}{lrrrrrrrr}
\toprule
& true &\multicolumn{2}{c}{\underline{\hspace{10mm} $n=200$ \hspace{7mm}}} &\multicolumn{2}{c}{\underline{\hspace{10mm} $n=400$\hspace{7mm}}} &\multicolumn{2}{c}{\underline{\hspace{10mm} $n=800$\hspace{7mm}}}\\
quantity & value & adjusted & unadjusted & adjusted & unadjusted & adjusted & unadjusted\\
\midrule
$\mbox{sd}(\hat \beta_{10})$ & 2.93 & 2.91 (0.28) & 3.41 (0.26) & 2.92 (0.20) & 3.41 (0.18) & 2.93 (0.14) & 3.41 (0.13) \\
$\mbox{sd}(\hat \beta_{11})$ & 0.10 & 0.10 (0.01) & 0.09 (0.01) & 0.10 (0.01) & 0.09 (0.00) & 0.10 (0.00) & 0.09 (0.00) \\
$\mbox{sd}(\hat \beta_{20})$ & 8.32 & 8.63 (1.68) & 8.67 (1.64) & 8.43 (1.11) & 8.45 (1.06) & 8.33 (0.78) & 8.35 (0.75) \\
$\mbox{sd}(\hat \beta_{21})$ & 0.16 & 0.16 (0.03) & 0.16 (0.03) & 0.16 (0.02) & 0.16 (0.02) & 0.16 (0.01) & 0.16 (0.01) \\
$\mbox{Cor}(\hat \beta_{10}, \hat \beta_{11})$ & -0.78 & -0.78 (0.03) & -0.81 (0.01) & -0.78 (0.02) & -0.81 (0.01) & -0.78 (0.01) & -0.83 (0.01) \\
$\mbox{Cor}(\hat \beta_{10}, \hat \beta_{20})$ & 0.60 & 0.59 (0.09) & 0 (0) & 0.60 (0.06) & 0 (0) & 0.60 (0.04) & 0 (0) \\
$\mbox{Cor}(\hat \beta_{10}, \hat \beta_{21})$ & -0.67 & -0.66 (0.08) & 0 (0) & -0.67 (0.05) & 0 (0) & -0.67 (0.03) & 0 (0) \\
$\mbox{Cor}(\hat \beta_{11}, \hat \beta_{20})$ & -0.38& -0.37 (0.09) & 0 (0) & -0.37 (0.06) & 0 (0) & -0.38 (0.04) & 0 (0) \\
$\mbox{Cor}(\hat \beta_{11}, \hat \beta_{21})$ & 0.63 & 0.63 (0.08) & 0 (0) & 0.63 (0.06) & 0 (0) & 0.63 (0.03) & 0 (0) \\
$\mbox{Cor}(\hat \beta_{20}, \hat \beta_{21})$ & -0.92 & -0.92 (0.02) & -0.92 (0.02) & -0.92 (0.02) & -0.92 (0.01) & -0.92 (0.01) & -0.92 (0.01) \\
\midrule
$\|\hat V_\beta - V_\beta\|_1$ & 0 & 28.3 & 40.5 & 17.7 & 30.7 & 12.5 & 26.5 \\
$\|\hat V_\beta^{-1} - V_\beta^{-1} \|_1$ \hspace{-10mm} & 0 & 418.1 & 1425.1 & 281.4 & 1555.0 & 194.4 & 1573.6 \\
\midrule
Type I & 1  & 1.9 & 3.8 & 1.4 & 3.7  & 0.9 &  3.1\\
errors & 5 & 7.2 & 11.3  & 6.0 & 10.1  & 5.1 & 9.1  \\
(\%)  & 10 & 13.2 & 17.1 & 11.3 & 15.4 & 9.7 & 13.4  \\
\bottomrule
\end{tabular}
\label{tab:2}
\end{table}

\subsection{Sorbinil eye trial simulations}
\label{se:sim2}
%Perhaps even more important than accurate, robust point estimation of the variance-covariance matrix is its use in making valid inferences on model parameters. 
We also examine the finite-sample accuracy of Result \ref{result:3} using simulations based on the Sorbinil eye trial from Example \ref{se:ex2}. To emulate the original dataset from \citet{RGL2006}, we first simulate values from Binomial distributions with 8 trials and then divide by 2, thereby generating data that take value only in $\{0, 0.5, 1, 1.5, \ldots, 3.5, 4\}$. Specifically, each data pair is given by $Y_L = 0.5 \sum_{j=1}^8 B_{Lj}$ and $
Y_R = 0.5 \sum_{j=1}^8 B_{Rj}$, where $(B_{Lj}, B_{Rj}), j=1,2,\ldots, 8,$ are correlated bivariate binary pairs generated from a random intercept and slope model,
\begin{eqnarray*}
B_{Lj}| \mbox{treatment},  \alpha_0, \alpha_1 &\sim& \mbox{Bernoulli}\left\{\mbox{expit}\left[0.303 + \alpha_0 + (-0.444 +\alpha_1) I(\mbox{sorbinil}_L = 1)\right]\right\} \\
B_{Rj} | \mbox{treatment},  \alpha_0, \alpha_1 &\sim& \mbox{Bernoulli}\left\{ \mbox{expit}\left[0.303 + \alpha_0 + (-0.444 + \alpha_1) I(\mbox{sorbinil}_R =1)\right]\right\} \ .
\end{eqnarray*}
%\begin{eqnarray*}
%Y_L| \mbox{treatment},  \alpha_0, \alpha_1 &\sim& \mbox{Binomial}\left\{8, \ \mbox{expit}\left[0.303 + \alpha_0 + (-0.444 +\alpha_1) I(\mbox{sorbinil}_L = 1)\right]\right\}/2 \\
%Y_R | \mbox{treatment},  \alpha_0, \alpha_1 &\sim& \mbox{Binomial}\left\{8, \  \mbox{expit}\left[0.303 + \alpha_0 + (-0.444 + \alpha_1) I(\mbox{sorbinil}_R =1)\right]\right\}/2
%\end{eqnarray*}
Here, the random intercept $\alpha_0 \sim N(0,0.2^2)$ is independent of the random slope $\alpha_1 \sim N(0, 0.2^2)$, and $\mbox{expit}[\cdot] = \exp(\cdot)/(1+\exp(\cdot))$. Thus, the baseline itchiness of 0.303 and treatment effect of sorbinil of 0.444 are the same for both left and right eyes, with correlation between eyes induced by the random effects terms $\alpha_0$ and $\alpha_1$. The parameter values were taken from a fitted model to the original dataset, so as to emulate the original values. We repeat this procedure to generate 5,000 datasets, each with $n=41$ observations, preserving the design of the original study.
 
We analyze each simulated dataset using the GEE model from Example \ref{se:ex2}, with 
%
%a pair of non-symmetric marginal logistic mean models,
%\begin{eqnarray*}
%E(Y_L/4|\mbox{ treatment} ) &\equiv& \mu_L = \mbox{expit}\left[\mu_{0} + \beta I(\mbox{sorbinil}_L =1) \right] \ , \\
%E(Y_R/4|\mbox{ treatment} ) &\equiv& \mu_R = \mbox{expit}\left[\mu_{0} + \mu_{0R} + (\beta + \beta_{R}) I(\mbox{sorbinil}_R =1)\right] \ .
%\end{eqnarray*} 
%We are then interested in testing for symmetry, i.e., we test the compound hypothesis $H_0: \mu_{0R} = \beta_{R}= 0$. 
%
%For the working variance-correlation structure, we use a pair of quasi-Binomial variance functions with constant correlation,
%\begin{eqnarray*}
%\mbox{Var}(Y_L/4| \mbox{treatment}) &=& \phi \, \mu_L \left( 1- \mu_L \right)\\
%\mbox{Var}(Y_R/4| \mbox{treatment}) &=& \phi \, \mu_R \left( 1- \mu_R \right)\\
%\mbox{Corr}(Y_L, Y_R | \mbox{treatment}) &=& \rho \ .
%\end{eqnarray*}
the working correlation either set to 0 exactly (working independence) or left unspecified in $(-1,1)$. We also consider the constant odds ratio model of \citet{Yee2015}, which characterises the dependence between individual bivariate binary responses $(B_L, B_R)$ via
$$
\frac{P(B_{L} =1, B_{R} =1) \,P(B_{L} =0, B_{R} =0)}{P(B_{L} =0, B_{R} =1) \, P(B_{L} =1, B_{R} =0)} = e^\gamma \ .
$$
This induces a within-vector correlation between the two eyes of the form
$$
\mbox{Corr}(B_L, B_R) = \frac{p_{11}(\gamma; \mu_L, \mu_R) - \mu_L \mu_R}{\sqrt{\mu_L(1-\mu_L) \, \mu_R(1-\mu_R)}} \ ,
$$
where $p_{11}(\gamma; \mu_L, \mu_R)$ is the solution to $
p_{11}(1-\mu_L-\mu_R + p_{11}) = (\mu_L -p_{11})(\mu_R - p_{11}) e^\gamma
$. Note that the working variance functions and the correlation structures are misspecified in all three GEE models considered here. It is in such situations where a sandwich adjustment may prove to be invaluable. In particular, for the third model, the approach considered in this note can be viewed as a ``plug-in" to the \texttt{vglm} function from the \texttt{VGAM} R package of \citet{Yee2015}, offering a robust post-fitting adjustment to the standard errors and variances in case the assumed model is misspecified. The code for performing this post-fitting adjustment is particularly simple, with sample code for a trivariate binary example provided in Appendix \ref{ap:plugin}.

The structure of the data leads us naturally to test for {\it symmetry} in the two eyes, i.e., $H_0: \beta_{L0} = \beta_{R0}$ and $\beta_{L1} = \beta_{R1}$. We can do this using the $\mathcal{F}$-test prescribed in Result \ref{result:3},
with $M$ being the $2 \times 4$ matrix of contrasts,
$$
M = \begin{pmatrix}
1 & 0 & -1 & 0 \\
0 & 1 & 0 & -1 
\end{pmatrix} \ ,
$$
$\delta = (0, 0)^T$, and $\hat V_\beta$ an estimate of the variance using either the unadjusted model-based variance estimator, or the sandwich-adjusted variance estimator from Result \ref{result:2} of Section \ref{se:estimation}. We compare this statistic to an $\mathcal{F}$ distribution with 2 and 37 degrees-of-freedoms.

The Type I errors from 5000 simulations for testing for symmetry are given in Table \ref{table:3}. We see that even for moderately small sample sizes of $n=41$ the accuracy of joint inferences show a significant improvement after the sandwich adjustment, regardless of the working correlation model. In contrast, the Type I errors using the naive variance estimator are consistently biased under model misspecifications, and are particularly sensitive to the working correlation.

Also displayed in Table \ref{table:3} are the Type I errors from a fully parametric Gaussian copula bivariate binary regression model, under both working independence and unspecified correlation models, using the R package \texttt{gcmr} \citep{MV2015}. The data-generating mechanism using a latent random effects term means that the copula model is also misspecified. To account for this, the \texttt{gcmr} package can be combined with the \texttt{sandwich} package \citep{Zeileis2006} which performs sandwich adjustments to the estimated variances. 

We see from Table \ref{table:3} that the Type I errors for the Gaussian copula bivariate binary model are more biased than for GEE approaches. While sandwich adjustments do improve the Type I errors, they only correct the bias to a limited degree. This highlights the sensitivity of inferences on parametric model assumptions, especially in finite-sample settings. From a computational point of view, the Gaussian copula bivariate binary model implemented via \texttt{gcmr} took over two orders of magnitude longer to fit than the GEE models. The average run time for \texttt{gcmr} with unspecified correlation was 18.82 seconds per simulated dataset, while the average run time was no more than 0.06 seconds for any of the GEE methods . Another limitation of the \texttt{gcmr} package is that it (currently) cannot handle response vectors of mixed type.

\begin{table}[H]
\footnotesize
\caption{Eye study simulations -- Type I errors (\%) for testing symmetry at nominal 10\%, 5\% and 1\% levels using GEE model with unadjusted and sandwich-adjusted estimates of variance, and Gaussian copula bivariate Binomial model. $N = 5000$ simulations each with sample size $n=41$.}
\centering
\begin{tabular}{cc rrr rrr}
\toprule
& & \multicolumn{3}{c}{\underline{unadjusted}} & \multicolumn{3}{c}{\underline{adjusted}} \\
method & working correlation  & 10\% & \ 5\% & \ 1\% & 10\% &\ 5\% & \ 1\% \\
\midrule
GEE & independence & 5.3 & 2.0 & 0.1 & 10.6 & 5.3 & 0.9 \\
& unspecified &  6.9 & 3.8 & 0.7 & 10.2 & 5.8 & 1.2 \\
& odds ratio & 7.4 & 3.2 & 0.5 & 8.9 & 4.1 & 0.8 \\
Copula & independence & 6.7 & 3.0 & 0.4 & 7.8 & 3.7 & 0.5 \\
& unspecified & 6.4 & 2.8 & 0.6 & 7.0 & 3.2 & 0.6 \\
\bottomrule
\end{tabular}
\label{table:3}
\end{table}

\vspace{-3mm}
\section{Data analysis examples}
\label{se:data}
\subsection{Burn injury data}
As mentioned in the introduction, there aren't many models that can handle continuous-binary response pairs. The method of \citet{FL1995} is useful when it is sensible to think of the continuous response being generated conditional on the binary response. However, here it is perhaps more tenable to think of incidence of death as being conditional on burn severity, and not vice versa. Unfortunately, flipping the order of conditioning leads to the marginal mean of the binary component being no longer logistic. In this sense, the hierarchical continuous-binary model of \citet{FL1995} is not exchangeable in the two vector components. 
%{\color{magenta} A lack of exchangeability is also exhibited by the joint probit-normal model of \citet[][Section 24.2.1]{MV2005}, whereby the binary component is dichotomized from an unobserved continuous latent variable by conditioning on the observed continuous component}. 
This is precisely when a GEE approach proves to be invaluable.

If we had observed the two variables separately, we would have fitted two independent models to the data. This corresponds to the GEE model from Example \ref{se:ex1} with the working independence correlation model. The sandwich correction from Result \ref{result:2} then allows us to adjust for within-subject correlations post-fitting.

The fitted marginal mean models using this GEE are
\begin{eqnarray*}
\hat E(\mbox{burn severity}\, |\, \mbox{age}) &=& 6.7118 + 0.0035 \ \mbox{age}\ ,\\
\hat P(\mbox{death}\, |\, \mbox{age}) &=& \frac{ \exp(-3.6891 + 0.0508 \ \mbox{age}) }{1+\exp(-3.6891 + 0.0508 \ \mbox{age}) } \ ,
\end{eqnarray*} 
with the adjusted standard errors of the slopes estimated as se$(\hat \beta_{11}) =0.0017$ and se$(\hat \beta_{21})= 0.0051$, respectively. We can conclude that age has significant marginal relationships with both burn severity and risk of death.

\begin{figure}[H]
\centering
\includegraphics[trim ={0 7mm 10mm 15mm}, clip, scale=0.7]{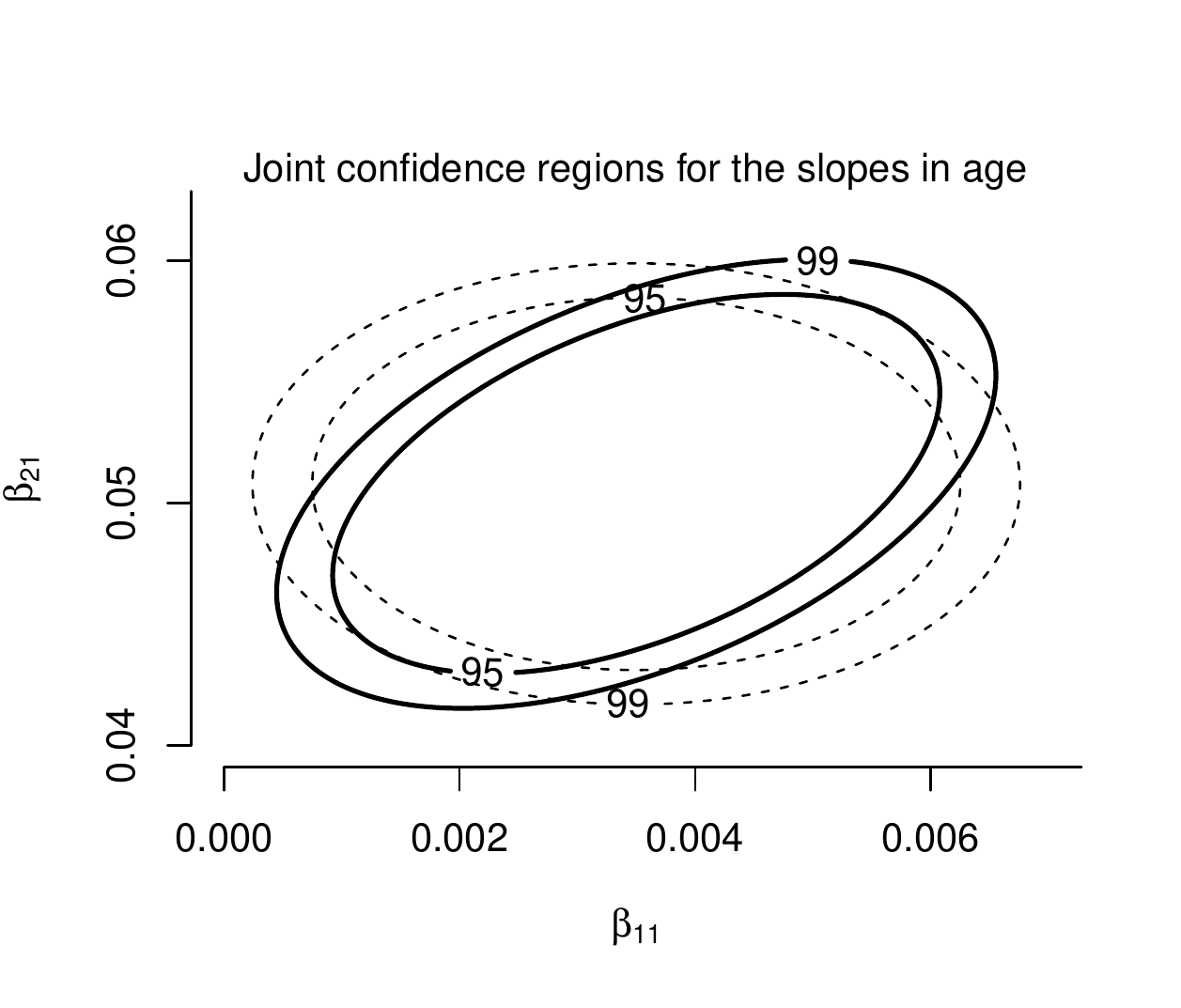}
\caption{Burn injury data -- joint 95\% and 99\% adjusted (solid) and unadjusted (dashed) confidence regions for the slopes in age.}
\label{fig:1}
\end{figure}

Perhaps more interesting than marginal effects is the joint effect of age on burn severity and incidence of death. By inverting the $\mathcal{F}$-test of Result \ref{result:3} with 
$$
M = \begin{pmatrix}
0 & 1 & 0 & 0 \\
0 & 0 & 0 & 1
\end{pmatrix} \ 
$$
and various values of $\delta$, we can obtain joint confidence regions for the two slopes $\beta_{11}$ and $\beta_{21}$. For example, the 95\% and 99\% joint confidence regions using the adjusted (solid) and unadjusted (dashed) variances are displayed in Figure \ref{fig:1}. In contrast to the unadjusted confidence regions, the adjusted regions place little confidence in the top left or bottom right quadrants, indicating that it is not plausible for age to have a strong relationship with burn severity without also having a strong relationship with incidence of death, or vice versa. It also suggests that the relationships of age with severity and with death are in the same direction.

%The estimated coefficients along with their naive and robust standard errors are displayed in Table 3.
%
%
%\begin{table}[H]
%\caption{Burn injury dataset: estimated coefficients, naive and robust standard errors and z-scores}
%\centering
%\begin{tabular}{ll|rrrr}
%Component & Coefficient & Estimate & Naive SE & Robust SE & Robust z \\
%\hline
%Burn area & Intercept & 6.7118 & 0.0690 & 0.0640 &  104.87 \\
%&  Age & 0.0035 & 0.0018 & 0.0017 & 2.06 \\
%Death & Intercept & -3.6891 & 0.2342 & 0.2643 &  -13.96 \\
%&  Age & 0.0508 & 0.0046 & 0.0050 &  10.16\\
%\end{tabular}
%\end{table}

\subsection{Sorbinil Retinopathy Trial data}

To analyze the sorbinil eye trial dataset from \citet{RGL2006}, we fit the GEE model from Example \ref{se:ex2} with an unspecified correlation. The nature of the data leads us to test whether the model can be simplified to one that is symmetric in the two eyes, that is, whether $\beta_{10} = \beta_{20}$ and $\beta_{11} = \beta_{21}$. Using Result \ref{result:3}, the $\mathcal{F}$ statistic for this test is 0.91 on 2 and 37 degrees of freedom, giving a p-value of 0.41. A symmetric model is therefore acceptable. 

In addition to symmetry considerations, the nature of the data also leads us to ask whether there may be an {\it interference} effect -- that is, does the treatment applied to one eye affect the response in the other eye? It is plausible to think that the application of treatment simultaneously to both eyes may have an effect on each eye that is different to the effect of treatment in isolation. 

To test for interference, we fit the expanded symmetric mean model,
\begin{eqnarray*}
E(\tilde Y_L|\mbox{ treatment} ) &=& = \frac{\exp \left[\beta_{0} + \beta_{1} I(\mbox{sorbinil}_L =1) + \beta_2 I(\mbox{sorbinil}_R = 1) \right]}{1+\exp \left[\beta_{0} + \beta_{1} I(\mbox{sorbinil}_L =1) + \beta_2 I(\mbox{sorbinil}_R = 1) \right] } \ , \\
E(\tilde Y_R|\mbox{ treatment} ) &=& = \frac{\exp \left[\beta_{0} + \beta_{1} I(\mbox{sorbinil}_R =1) + \beta_2 I(\mbox{sorbinil}_L = 1) \right]}{1+\exp \left[\beta_{0} + \beta_{1} I(\mbox{sorbinil}_R =1) + \beta_2 I(\mbox{sorbinil}_L = 1) \right] }  \ .
\end{eqnarray*} 
The additional parameter $\beta_2$ measures the interference effect. For this dataset, the estimated interference effect is $\hat \beta_2 = 0.018$ with an adjusted standard error of $\hat{\mbox{sd}}(\hat \beta_2)=0.162$. There is no evidence of interference here.

The final fitted model is summarized in Table \ref{table:4}. We see that sorbinil treatment is associated with an estimated reduction in the itchiness score from $4 \times \mbox{expit}(0.303) = 2.30$ to $4 \times \mbox{expit}(0.303 - 0.444) = 1.86$ on the original scale. This reduction is highly significant, with an associated p-value of $P(T_{39} \le -3.42) = 7.5 \times 10^{-4}$.

\begin{table}[H]
\footnotesize
\caption{Sorbinil eye trial dataset -- estimated coefficients, naive and adjusted standard errors and z-scores}
\centering
\begin{tabular}{lrrrr}
\toprule
Coefficient & Estimate & Naive SE & Adjusted SE & Adjusted z \\
\midrule
Intercept & 0.303 & 0.129 & 0.103 & 2.95 \\
Sorbinil & -0.444 & 0.144 & 0.130 & -3.42 \\
\bottomrule
\end{tabular}
\label{table:4}
\end{table}

%\begin{table}[H]
%\caption{Sorbinil eye trial dataset -- estimated coefficients, naive and robust standard errors and z-scores}
%\centering
%\begin{tabular}{l|rrrr}
%Coefficient & Estimate & Naive SE & Robust SE & Robust z \\
%\hline
%Intercept & 0.288 & 0.219 & 0.166 & 1.73 \\
%Sorbinil & -0.431 & 0.209 & 0.164 & -2.63 \\
%Interference & 0.018 & 0.209 & 0.162 & 0.11
%\end{tabular}
%\end{table}

\section{Further topics}
\label{se:further}
\subsection{Higher dimensional responses}
The connection between GEEs and general vector regression made in this paper extends beyond the bivariate cases considered here. The framework is particularly useful for response vectors with three or more components, where specification of appropriate joint distributions for the data may be even more difficult. For response vectors with components of the same type, the results of this note can be implemented as a ``plug-in" to existing methods, such as the \texttt{vglm} function in the \texttt{VGAM} package of \citet{Yee2015, Yee2016}. 

For example, \citet{Yee2016} considers the presence/absence of three plant species (cyadea, beitaw and kniexc) at 392 locations in the Hunua ranges in Auckland as a vector regression problem, with each response being a trivariate binary random vector and the predictor variable being altitude. A set of logistic mean models is specified for each component,
$$
P(Y_k = 1 | \mbox{altitude}) = \mu_k = \frac{\exp \left(\beta_{k0} + \beta_{k1} \mbox{altitude}\right)}{1+ \exp \left(\beta_{k0} + \beta_{k1} \mbox{altitude}\right)} \ , \quad k \in \{\mbox{cyadea, beitaw, kniexc}\} \ ,
$$
with a corresponding set of variance functions given by
\begin{equation}
\label{eq:varhunua}
\mbox{Var}(Y_k| \mbox{altitude}) = \mu_k \, (1-\mu_k) \ .
\end{equation}
Dependence between plant species at each location is modelled via a constant log-odds ratio model,
\begin{equation}
\label{eq:logoddshunua}
\frac{P(Y_k = 1, Y_{k'} = 1)P(Y_k = 0, Y_{k'} = 0)}{P(Y_k = 0, Y_{k'} = 1P(Y_k = 1, Y_{k'} = 0)} = e^{\gamma_{kk'}} \ , \quad k \ne k' \ .
\end{equation}
This induces within-vector correlations that depend on the marginal means and the parameters $\{\gamma_{kk'}\}$.

The model is then fit using the \texttt{vglm} function from the \texttt{VGAM} R-package. The estimated regression coefficients and standard errors are given in Table \ref{table:5}. These assume that the variance (\ref{eq:varhunua}) and dependence (\ref{eq:logoddshunua}) models are correctly specified. To account for possible model misspecification, the sandwich estimator from Result \ref{result:2} can be implemented as a ``plug-in" to \texttt{vglm}. The code for this post-fitting adjustment is particularly simple and is provided in Appendix \ref{ap:plugin}. The corresponding adjusted standard errors and $z$ statistics are given in columns 5 and 6 of Table \ref{table:5}, respectively. The overall similarity between the adjusted and unadjusted standard errors suggests that the working variance and dependence models are reasonable in this case. However, the adjusted standard errors are always asymptotically valid regardless of the true underlying variance-correlation model.

\begin{table}
\centering
\footnotesize
\caption{Presence/absence of three plant species in the Hunua ranges -- estimated coefficients, unadjusted and adjusted standard errors, and adjusted $z$-scores}
\begin{tabular}{llrrrr}
\toprule
Component & Coefficient & Estimate & Naive SE & Adjusted SE & Adjusted $z$ \\
\midrule
Cyadea & Intercept & -0.977 & 0.222 & 0.170 & -5.74\\
 & Altitude ($\times 10^3$) & -0.570 & 0.921 & 0.813 & -0.70\\
Beitaw & Intercept & -1.890 & 0.265 & 0.181 & -10.47 \\
& Altitude ($\times 10^3$)& 3.850  & 0.962 & 0.870 & 4.42\\
Kniexc & Intercept & -0.377 & 0.197 & 0.197 & -1.91\\
& Altitude ($\times 10^3$) & 1.611 & 0.970 & 1.188 & 1.36 \\
\bottomrule
\end{tabular}
\label{table:5}
\end{table}

\subsection{Missing components}
\label{se:missing}
The GEE framework for vector regression is readily adaptable to handle cases where some components are missing completely at random for some observations. Suppose, for example, that component $k$ is missing from observation $i$. Then, the contribution of the $i$th observation to the estimating equation can be modified to $
D_{i,-k} W_{i, -k}^{-1} S_{i, -k}
$, where $D_{i,-k}$ and $W_{i, -k}$ denote the matrices $D_i$ and $W_i$, respectively, with row and column $k$ removed, and $S_{i,-k}$ is the vector $S_i$ with element $k$ removed.

{\color{black} For responses that are missing at random, \citet{RRZ1995} developed a class of inverse-probability weighted GEE models for longitudinal data. This was extended by \citet{RRS1998} to handle nonignorable missingness. Adapting these methods to vector regression settings is a topic for future research.}

\subsection{A comparison with joint models}
A reviewer pointed out that there exists a multitude of joint models for repeated measures and longitudinal data \citep[e.g.,][Chapter 24]{HDD2000,MV2005}, some of which may be adaptable for vector regression problems. 
%have been successfully adapted to handle specific vector regression problems. 
However, the general applicability of such methods to vector regression settings is not immediate. For example, the model from \citet[][Chapter 24.2.1]{MV2005} dichotomizes one component of a bivariate normal vector to generate a probit--normal joint model for binary-continuous response pairs, but it is not clear how this dichotomization can be generalized to a discretization for handling count--continuous responses. Moreover, even if an analogous discretization can be formulated, the resulting mean model will generally not be of an easily-interpretable form, such as a log-linear model. In contrast, model specification in the GEE vector regression approach is done simply via the first two moments, so that count components can be handled by specifying, say, a log-linear mean model in (\ref{eq:mean}) and a linear variance function in (\ref{eq:var}); this is no harder to specify than for continuous or binary components.

A more generic joint modelling approach is the generalized linear mixed model (GLMM) formulation \citep[e.g.,][Chapter 24.2.3]{MV2005} which induces within-vector correlations via subject-specific random effects. While mixed models for response vectors of mixed type are conceptually no different to mixed models for repeated measures of the same type, on a practical level it is well-known that for non-Gaussian responses misspecification of the random effects distribution can lead to severely biased estimation and inferences on model parameters \citep[e.g.,][]{LAM2008}. Moreover, mixed models generally do not preserve marginal mean structures when non-identity links are used. For example, including random effects in a binary logistic model typically leads to a marginal mean model that is no longer logistic. In contrast, GEE vector regression explicitly models the marginal means via (\ref{eq:mean}), so that model specification and parameter interpretation are both simple and invariant to the underlying dependence structure.

\citet{HDD2000} formulated a class of joint models for a sequence of longitudinal measures and an associated sequence of event times. This class covers the models of \citet{TDW1995} and \citet{WT1997} as special cases. The association between the two sequences of observations is characterized through a latent bivariate Gaussian process indexed by time. If there is no longitudinal aspect, this approach reduces to a GLMM with normal random intercepts, and so the drawbacks of mixed models also hold here. Additionally, inferences on model parameters assume that the underlying latent Gaussian process is correctly specified, and (currently) no adjustments are available in the case of model misspecification. However, one very attractive feature of the \citet{HDD2000} approach is that it can handle within-subject dependence both between the bivariate vector components and across time. The extension of the GEE vector regression approach to allow for longitudinal dependence will be an important and invaluable development.

\section{Discussion}
The idea of extending GEEs to general vector regression problems has been suggested in the literature, but generally in an ad hoc manner. This paper formalises the connection.

 %The extended GEE vector regression approach is particularly useful when a full distributional model for the response vector is hard to specify. 
%This may be because the vector components are of mixed type %, such as the binary-continuous pairs from Example \ref{se:ex1}, 
%or because the joint sample space is somewhat tricky
%, such as the bivariate lattice for pairwise itchiness scores from Example \ref{se:ex2}. 
The main attraction of the GEE approach 
%in such scenarios
is its simplicity in model specification and the guarantee of asymptotically correct inferences without requiring correct models for the variances, correlations, or underlying joint distributions for the data. This is particularly useful when a full distributional model for the response vectors is hard to specify, such as in Examples \ref{se:ex1} and \ref{se:ex2}. 

The guarantee of asymptotically valid inferences relies on the sandwich estimator of variance (Result \ref{result:2}). 
%The GEE vector regression approach relies on the sandwich estimator of variance (Result \ref{result:2}) for asymptotically valid inferences on model parameters 
%without requiring correct specification of the variances, correlations, or underlying joint distributions for the data. 
Although our simulations demonstrate that GEE vector regression can be more robust to model misspecification than copula-based methods 
for moderate sample sizes, it is well-known that the sandwich estimator can be severely biased in small-sample settings.
%However, it is well-known that the sandwich estimator can be severely biased in small-sample settings. 
Small-sample modifications of the sandwich estimator have recently been proposed for longitudinal settings \citep[e.g.,][]{FG2001, MD2001}. Extending these modifications to the general vector regression case is a direction for future research.

R software for implementing general GEE vector regression models is currently in development. One arm of the development is to implement ``plug-ins" to the many multivariate families available in \texttt{vglm} from the \texttt{VGAM} package \citep{Yee2016}. This will be useful for vector responses of the same type. A second arm will implement GEE models for vectors of mixed type, with scope for handling a wide range of working correlation structures. Although a general software package is not yet available, GEE vector regression models are straightforward to implement on a case-by-case basis, as demonstrated via the simple R routines for implementing Examples \ref{se:ex1} and \ref{se:ex2} provided in the Appendix.

\appendix

\section{Sorbinil trial dataset}
\label{se:appA}
Table 1 goes here.

\section{R Codes}
\subsection{GEE for continuous-binary pairs with working independence}
\label{ap:B1}
\footnotesize
\begin{verbatim}
# fit independent models for each component
fit1 = glm(y1~X1)
fit2 = glm(y2~X2, family = quasibinomial)

# fitted values
mu1 = fit1$fitted
mu2 = fit2$fitted

# working variances
W1 = rep(summary(fit1)$disp, n)
W2 = summary(fit2)$disp*mu2*(1-mu2)

# model-based variance-covariance matrix
naive.vcov =  bdiag(summary(fit1)$cov.scaled, summary(fit2)$cov.scaled)

# compute sandwich estimator of variance
meat = 0
for (i in 1:n){
   Di = bdiag(model.matrix(fit1)[i,],(mu2[i]-mu2[i]^2)*model.matrix(fit2)[i,])
   Wi.inv = diag(c(1/W1[i], 1/W2[i]))
   Si = rbind(y1[i] - mu1[i], y2[i]-mu2[i])
   meat = meat + Di%*%Wi.inv%*%Si%*%t(Si)%*%Wi.inv%*%t(Di)
}
sandwich.vcov = naive.cov%*%meat%*%naive.cov
\end{verbatim}

%\subsection{A ``plug-in" to \texttt{vglm}}
%\begin{verbatim}
%library(VGAM); library(Matrix)
%
%# fit bivariate binary model to sorbinil eye trial data using vglm
%fit = vglm(cbind(BL, BR) ~ sorbinil, family=binom2.or, 
%      xij = list(sorbinil ~ sorbinil.l + sorbinil.r + fill1(sorbinil.l)), 
%      form2 = ~ sorbinil + sorbinil.l + sorbinil.r + fill1(sorbinil.l))
%	
%naive.vcov = vcov(fit)[c(1,4,2,5), c(1,4,2,5)]
%# note that the order of terms outputted from vglm is:
%# intercept1, intercept2, gamma, slope11, slope21, slope12, slope22,...
%# which differs from our order, which is:
%# intercept1, slope11, slope12,..., intercept2, slope21, slope22,...
%
%# fitted marginal probabilities
%mu1 = apply(fitted(fit)[,c(3,4)], 1, sum)
%mu2 = apply(fitted(fit)[,c(2,4)], 1, sum)
%
%# sandwich estimator of variance
%meat = 0 ;
%bread = 0
%for (i in n){
%   p11 = fitted(fit)[i,4]
%   Wi = rbind(c(mu1[i]*(1-mu1[i]),  p11-mu2[i]*mu2[i]),
%              c(p11-mu1[i]*mu2[i], mu2[i]*(1-mu2[i])))
%   Di = bdiag(mu1[i]*(1-mu1[i])*cbind(1, sorbinil.l[i]),
%              mu2[i]*(1-mu2[i])*cbind(1, sorbinil.r[i]))
%   Si = rbind(y1[i] - mu1[i], y2[i] - mu2[i])
%   meat = meat + t(Di)%*%solve(Wi)%*%Si%*%t(Si)%*%solve(Wi)%*%Di
%   bread = bread + t(Di)%*%solve(Wi)%*%Di
%}
%sandwich.vcov = solve(bread)%*%(meat/8)%*%solve(bread)
%\end{verbatim}

\subsection{A ``plug-in" to \texttt{vglm}}
\label{ap:plugin}
\begin{verbatim}
library(Matrix); library(VGAM); attach(hunua);

# fit trivariate binary regression model to Hunua dataset
fit <- vglm(cbind(cyadea, beitaw, kniexc) ~ altitude, loglinb3)

# fitted probabilities for each outcome
fitted = fitted(fit)

# fitted marginal probabilities
mu.c = apply(fitted[,c(5,6,7,8)], 1, sum)
mu.b = apply(fitted[,c(3,4,7,8)], 1, sum)
mu.k = apply(fitted[,c(2,4,6,8)], 1, sum)
mu.cb = apply(fitted[,c(7,8)], 1, sum)
mu.ck = apply(fitted[,c(6,8)], 1, sum)
mu.bk = apply(fitted[,c(4,8)], 1, sum)

# fitted variances and covariances
var.cc = mu.c*(1-mu.c);   var.bb = mu.b*(1-mu.b);   var.kk = mu.k*(1-mu.k) ;
cov.cb = mu.cb-mu.c*mu.b; cov.ck = mu.ck-mu.c*mu.k; cov.bk = mu.bk-mu.b*mu.k;

# compute sandwich estimator of variance	
meat = 0 ;
bread = 0
for (i in 1:392){
 Wi = rbind(c(var.cc[i], cov.cb[i], cov.ck[i]), 
            c(cov.cb[i], var.bb[i], cov.bk[i]), 
            c(cov.ck[i], cov.bk[i], var.kk[i]))
 Di = bdiag(mu.c[i]*(1-mu.c[i])*cbind(1, altitude[i]), 
            mu.b[i]*(1-mu.b[i])*cbind(1, altitude[i]),  
            mu.k[i]*(1-mu.k[i])*cbind(1, altitude[i]))
 Si = rbind(cyadea[i] - mu.c[i], beitaw[i] - mu.b[i], kniexc[i] - mu.k[i])
 meat = meat + t(Di)%*%solve(Wi)%*%Si%*%t(Si)%*%solve(Wi)%*%Di
 bread = bread + t(Di)%*%solve(Wi)%*%Di
}
sandwich.vcov = solve(bread)%*%(meat)%*%solve(bread)
\end{verbatim}

\end{document}